\documentclass[preprint,review,12pt]{elsarticle}

\journal{Chemical Engineering Journal}
\usepackage{amssymb}
\usepackage{amsmath}
\usepackage{graphicx}
\usepackage{dcolumn} % Align table columns on decimal point
\usepackage{subcaption}
\usepackage{bm}% bold math

\usepackage[version = 4]{mhchem} % Formulas using \ce{}
\usepackage{siunitx}
\usepackage{hyperref}
\usepackage[nameinlink]{cleveref}
\crefname{table}{Table}{Tables}
\Crefname{table}{Table}{Tables}
\crefname{figure}{Fig.}{Figs.}
\Crefname{figure}{Figure}{Figures}
\crefname{equation}{Eq.}{Eqs.}
\Crefname{equation}{Equation}{Equation}
\crefname{section}{Sec.}{Secs.}
\Crefname{section}{Section}{Sections}

\begin{document}

\begin{frontmatter}

\title{Density diversity in training data governs thermodynamic transferability of machine learning interatomic potentials}

\author[snu]{Minwoo Kim}
\author[snu]{Seungtae Kim}
\author[snu]{Je-Yeon Jung}
\author[khu]{Min Young Ha\corref{cor1}}
\ead{myha@khu.ac.kr}
\author[snu]{Won Bo Lee\corref{cor2}}
\ead{wblee@snu.ac.kr}

\cortext[cor1]{Corresponding author}
\cortext[cor2]{Corresponding author}

\address[snu]{Department of Chemical and Biological Engineering, Institute of Chemical Processes, Seoul National University, Seoul 08826, Republic of Korea}
\address[khu]{Department of Chemical Engineering, Kyung Hee University, Yongin, 17104, Republic of Korea}

\begin{abstract}
Machine learning interatomic potentials (MLIPs) offer first-principles accuracy with reduced computational cost, but their transferability across different thermodynamic states remains questionable, particularly for fluid systems where molecules experience local environments far from crystalline equilibrium. Here, we demonstrate that diversifying the density of training configurations, rather than temperature, is the most effective strategy for building thermodynamically transferable MLIPs within a fixed computational budget. We first show that foundation MLIPs trained on solid-state databases accurately describe liquid-like densities but fail at gas-like conditions, while molecular-database-trained models exhibit the opposite behavior. Controlled from-scratch training and distillation experiments confirm that density-diverse datasets resolve both failure modes, whereas temperature-diverse datasets cannot compensate for missing density regimes. Coordination number analysis reveals the physical origin of this behavior: local coordination topology is more susceptible to density than temperature, leading to further structural diversity. These results establish density diversity as a design principle for thermodynamically transferable MLIPs and provide a validation framework for assessing the thermodynamic coverage of both foundation and from-scratch models, enabling reliable atomistic simulation of fluid-phase processes across diverse operating conditions.
\end{abstract}


\begin{highlights}
\item MLIPs systematically fail outside the density regime of their training data.
\item Density-diverse training broadens MLIP coverage from gas to dense liquid states.
\item Distillation with density-diverse data yields fast, broadly transferable MLIPs.
\item Coordination number serves as a physical descriptor of MLIP thermodynamic coverage.
\item Density variation alters coordination topology more than temperature variation.
\end{highlights}

\begin{graphicalabstract}
\includegraphics{"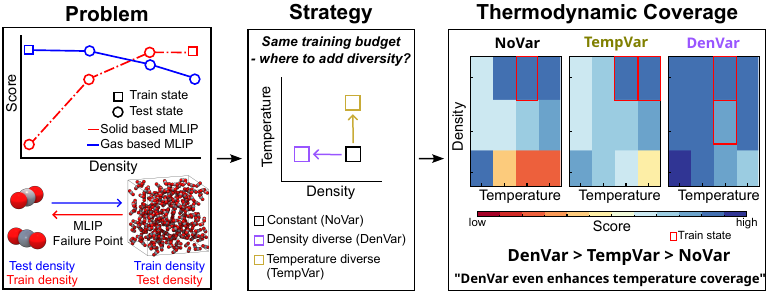"}
\end{graphicalabstract}

\end{frontmatter}

% \linenumbers
\section{Introduction}
Many fluid-phase operations central to chemical engineering, including supercritical extraction\cite{simeski2023supercritical,yoon2017molecular,li2025molecular,han2024understanding}, electrolyte transport\cite{muralidharan2018molecular,park2025lithium,kubisiak2017molecular}, and reactive flows in non-ideal solvents\cite{winetrout2024implementing,mao2023classical}, require consistent fidelity across a wide range of temperature, pressure, and density. Molecular dynamics (MD) simulations resolve such phenomena at the atomic scale, with their accuracy hinging on the underlying potential energy surface (PES). First-principles methods offer high accuracy, but their computational cost confines them to small-scale or short-timescale simulations far from process-relevant length and time scales. Machine learning interatomic potentials (MLIPs) bridge this gap by learning the PES from quantum mechanical calculations, enabling simulations with near first-principles accuracy at a fraction of the cost. These MLIP-based approaches have accelerated atomistic studies across materials science, chemistry, and chemical engineering, enabling property and reactivity predictions in systems previously inaccessible to ab initio MD.\cite{choi2025reliable,zhang2026unraveling,unke2021machine,yang2022using,avula2023understanding,kim2024anomalous,ryu2024understanding,ryu2025exploring,yu2024spatiotemporal}

Two complementary strategies have emerged for constructing MLIPs: From-scratch training on system-specific data and foundation models trained on large-scale, chemically diverse databases. Researchers have developed from-scratch MLIPs using architectures such as DeePMD-kit\cite{zeng2025deepmd}, NequIP\cite{batzner20223}, MACE\cite{Batatia2022mace}, and SevenNet\cite{park_scalable_2024} to address physicochemical and process-relevant problems in diverse systems. These from-scratch models, trained specifically for particular material systems, demonstrate excellent predictive performance within their training domains. However, developing from-scratch MLIPs requires substantial first-principles data and researcher intuition to construct representative training sets. The cost of generating sufficient training data and the difficulty of sampling relevant configurational spaces confine from-scratch models to narrowly defined chemical and thermodynamic coverages, limiting their ability to span the operating envelopes encountered in process simulation.

Foundation MLIPs including CHGNet\cite{deng2023chgnet}, MACE-MP-0\cite{batatia2025foundation}, SevenNet-0\cite{kim_sevennet_mf_2024}, and SO3LR\cite{kabylda2025molecular} have emerged to extend chemical coverage across most elements in the periodic table. These models leverage graph neural networks and train on massive density functional theory (DFT) datasets such as the Materials Project\cite{Deng2023}, Alexandria Database\cite{ghahremanpour2018alexandria}, and Open Materials 2024\cite{barroso2024open}. Foundation MLIPs demonstrate remarkable transferability to solids and ionic systems without fine-tuning, occasionally extending their applicability even to systems outside their training domain. This success suggests a path toward universal PES applicable across broad chemical spaces, enabling large-scale simulations. However, the apparent transferability of foundation MLIPs in crystalline materials arises because their configurations occupy a narrow range of densities and near-equilibrium coordination environments.

Foundation models, trained predominantly on crystal-structure databases, capture the PES features relevant to these near-equilibrium configurations. Fluid-phase systems central to chemical engineering present a distinct challenge, as molecules explore broad configurational and thermodynamic envelopes that span gas-, liquid-, and supercritical states.\cite{yoon2017molecular,ha2018widom,ryu2025exploring} Across these regimes, transport, structural, and reactive properties shift sharply with thermodynamic state, requiring MLIPs to accurately describe configurations far from crystalline equilibrium. Recent benchmarks on ionic liquids and battery electrolytes have shown that foundation MLIPs without task-specific fine-tuning can reproduce qualitative structural features such as radial distribution functions, yet exhibit non-negligible deviations in more sensitive properties including dihedral angle distributions, density, diffusivity, and ionic conductivity, quantities that directly govern process design and performance.\cite{batatia2025foundation,park2025ionic,grunert2025modeling,ju2025application,goodwin2024transferability}

Both from-scratch and foundation MLIPs face a common challenge of data efficiency, as expanding training datasets increases computational cost and slows training without guaranteeing improved transferability. For both paradigms, researchers confront a persistent question of practical importance for process simulation: How should training datasets be constructed to maximize thermodynamic transferability within a fixed training budget, so that a single model spans the temperature, pressure, and density envelope of a target operation? Current practices rely on researcher intuition, and there is no systematic guideline for selecting training configurations across thermodynamic variables including temperature, density, and composition. While previous studies have proposed training at elevated temperatures to sample higher-energy configurations, the role of density diversity remains unexplored.\cite{stark2024benchmarking}

In this work, we first assess thermodynamic coverage of representative foundation MLIPs across gas- to liquid-like densities, revealing complementary failure modes that map directly onto the density regimes of their training databases (\cref{sec:1,sec:2}). We then conduct controlled from-scratch training and distillation experiments to isolate the roles of density and temperature diversity, establishing that density-diverse training achieves the broadest thermodynamic transferability within a fixed computational budget (\cref{sec:3,sec:4}). Finally, coordination number analysis is used to elucidate the physical mechanism underlying this hierarchy, demonstrating that density variation alters the local coordination environment far more effectively than temperature variation (\cref{sec:5}).

\section{Results and discussion}
\begin{table*}
    \centering
    % \begin{ruledtabular}
    \begin{tabular}{c|c|c|c|c}
    \hline
    & \multicolumn{4}{c}{Database} \\
    \hline
    Model&MPTrj&OMat24&MPTrj+sAlex+OMat24&MD DB\\
    \hline
    7net&7net-0&7net-OMAT&7net-MF-OMPA&-\\
    \hline
    SO3krates&-&-&-&SO3LR\\
    \hline
    \end{tabular}
    % \end{ruledtabular}
    \caption{\label{tbl:1} Foundation MLIPs and their training database. MPTrj, sAlex, OMat stands for Materials Project Trajectory\cite{Deng2023}, subsampled Alexandria\cite{ghahremanpour2018alexandria}, and Meta Open Materials 2024 databases\cite{barroso2024open}, respectively. MD DB denotes SO3LR database which contains data recalculated with the PBE0-MBD exchange-correlation functional from GEMS\cite{unke2024biomolecular}, QM7-x\cite{hoja2021qm7}, AQM\cite{medrano2024dataset}, SPICE\cite{eastman2023spice}, and DES15k\cite{donchev2021quantum}.}
\end{table*}
To investigate the thermodynamic coverage of MLIPs as a function of the composition of the training database, we evaluated several representative foundation MLIPs, selected to span different training-database compositions. The training databases considered in this study are Materials Project Trajectory (MPTrj), Meta Open Materials 2024 (OMat24), Alexandria subsampled (sAlex), and MD DB~(\cref{tbl:1}). MPTrj, OMat24, and sAlex are inorganic solid-state databases computed with the generalized gradient approximation (GGA) and GGA+U exchange-correlation functionals.\cite{perdew1996generalized, anisimov1991band} MD DB was constructed by recalculating molecular-scale and partially solvated configurations at the Perdew-Burke-Ernzerhof hybrid (PBE0)\cite{adamo1999accurate} level with many-body dispersion (MBD) correction\cite{tkatchenko2012accurate}, and comprises GEMS fragments\cite{unke2024biomolecular}, 1M QM7-X molecules\cite{hoja2021qm7}, 60k AQM gas-phase molecules\cite{medrano2024dataset}, 33k SPICE dipeptides\cite{eastman2023spice}, and 15k DES molecular dimers.\cite{donchev2021quantum} These foundation models were developed to ensure broad chemical coverage by training on most elements across the periodic table.
Among the available architectures, we selected models based on the Scalable Equivariance-Enabled Neural Network (SevenNet, 7net)\cite{park_scalable_2024} and SO3krates.\cite{frank2022so3krates} Specifically, we examined 7net-0 trained on MPTrj, 7net-OMAT trained on OMat24, and 7net-MF-OMPA trained simultaneously on MPTrj, sAlex, and OMat24 through multi-fidelity learning. For all simulations with these models, Grimme's D3 dispersion correction with Becke-Johnson (BJ) damping was applied.\cite{grimme2010consistent, grimme2011effect} In addition, we included SO3LR,\cite{unke2024biomolecular} a SO3krates-based foundation MLIP developed for biomolecular simulation. SO3LR has shown high precision for small biomolecules, water, proteins, glycoproteins, and lipid bilayers, and was selected to compare the performance of a model trained on molecular-level data against those trained on solid-state databases.

\subsection{\label{sec:1}Assessment of Thermodynamic Coverage in Fluidic Systems}

\begin{figure}
    \centering
    \includegraphics[width=\columnwidth]{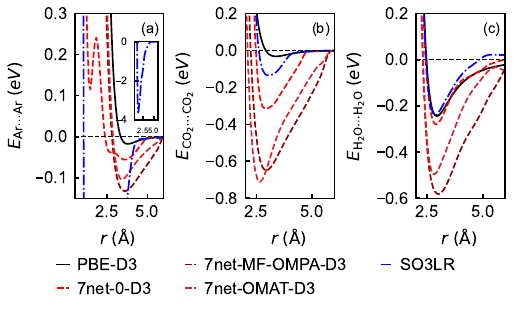}
    \caption{\label{fig:1}Pairwise interaction energy profiles for molecular dimers. (a) \ce{Ar-Ar}, (b) \ce{CO2-CO2}, (c) \ce{H2O-H2O} interaction energies as a function of interatomic distance computed with PBE-D3(BJ) (solid line, black), 7net-based foundation MLIPs (dashed lines: 7net-0-D3(light red), 7net-OMAT-D3(red), 7net-MF-OMPA-D3(dark red)), and SO3krates based foundation MLIPs (dash-dot: SO3LR (blue)). Dimer geometries used to generate each curve are provided in Fig.~S1 of the Supplementary Material.}  
    \phantomsubcaption\label{fig:1a}
    \phantomsubcaption\label{fig:1b}
    \phantomsubcaption\label{fig:1c}
\end{figure}

We begin our assessment of MLIP thermodynamic transferability by examining pairwise interaction energy profiles for molecular dimers. \cref{fig:1} presents interaction energy curves for \ce{Ar-Ar}, \ce{CO2-CO2}, and \ce{H2O-H2O} dimers as a function of intermolecular distance, with the corresponding well depths and equilibrium separations summarized in Tables~S1--S3 of the Supplementary Material. All foundation models were evaluated against PBE-D3(BJ) as reference. Interestingly, although the chemical complexity of intermolecular interactions increases in the order of Ar, CO2, and H2O, the accuracy of foundation MLIPs exhibits an inverse correlation, implying that the transferability limitations of foundation MLIPs arise not from insufficient model capability but from the coverage of the training data.

The \ce{Ar-Ar} interaction (\cref{fig:1a}) reveals significant discrepancies across all foundation MLIPs. The reference PBE-D3(BJ) calculation predicts an energy minimum of $-0.018$ \unit{eV} at $3.773$ \unit{\AA}. 7net-0-D3 and 7net-OMAT-D3 predict minima at $3.588$ \unit{\AA} with well depths of $-0.056$ \unit{eV} and $-0.104$ \unit{eV}, respectively, showing the smallest positional deviations among all models tested. 7net-MF-OMPA-D3 exhibits an even deeper well of $-0.133$ \unit{eV}. SO3LR shows the largest deviation, placing the energy minimum near $1.370$ \unit{\AA} with a depth of $-3.597$ \unit{eV}, indicating quantitatively incorrect behavior. 7net-0-D3 displays a double-well structure, rendering it qualitatively unsuitable for Ar simulations despite having the smallest minimum energy error among the 7net variants.

For the \ce{CO2} dimer (\cref{fig:1b}), two \ce{CO2} molecules were positioned in a crossed configuration with a 90\unit{\degree} angle between their molecular axes, and the interaction energy was computed as a function of inter-carbon distance. The reference PBE-D3(BJ) calculation gives an energy minimum of $-0.032$ \unit{eV} at $3.340$ \unit{\AA}. All foundation MLIPs reproduce a single energy minimum without the double-well artifacts observed in the \ce{Ar} case. SO3LR achieves the closest agreement with the reference, predicting a minimum at $2.983$ \unit{\AA} with a well depth of $-0.135$ \unit{eV}. 7net-0-D3 predicts $-0.316$ \unit{eV} at $2.874$ \unit{\AA}, while 7net-MF-OMPA-D3 and 7net-OMAT-D3 show larger deviations with well depths of $-0.649$ \unit{eV} and $-0.712$ \unit{eV}, respectively. Although qualitative agreement improves relative to the \ce{Ar} case, all models quantitatively overestimate the interaction strength.

The \ce{H2O} dimer (\cref{fig:1c}) demonstrates the best overall agreement among the three systems tested. Two water molecules were arranged in a hydrogen-bonding geometry with the donor \ce{O-H} bond aligned toward the acceptor oxygen lone pair, and the interaction energy was computed as a function of inter-oxygen distance. The reference PBE-D3(BJ) predicts a minimum of $-0.245$ \unit{eV}. 7net-0-D3 and SO3LR predict $-0.280$ \unit{eV} and $-0.239$ \unit{eV}, respectively, both within $0.04$ \unit{eV} of the reference. All foundation models reproduce the minimum position within $0.1$ \unit{\AA} of the reference value.

The progression from \ce{Ar} to \ce{CO2} to \ce{H2O} represents increasing complexity in the underlying intermolecular interactions, spanning Lennard-Jones, quadrupole, and hydrogen-bonding interactions, respectively. If the observed errors originated from insufficient model expressivity, one would expect accuracy to degrade as interaction complexity increases. However, \cref{fig:1} shows the opposite trend. Model accuracy improves from \ce{Ar} to \ce{H2O}, and the inter-model variance also decreases. This trend can be understood by examining the composition of the training databases. Systems of primary interest in materials and molecular science, such as water and organic molecules containing \ce{O}, \ce{C}, and \ce{H}, are well represented in the training data. In contrast, noble gases such as Ar are rarely included as target systems. This interpretation is consistent with element-resolved errors reported in MatBench\cite{dunn2020benchmarking}, where the prediction error increases in the order of \ce{O}, \ce{C}, and \ce{Ar}. These results indicate that the dominant source of error in pairwise interaction predictions is not model architecture but training data composition.

\begin{figure}
    \centering
    \includegraphics[width=\columnwidth]{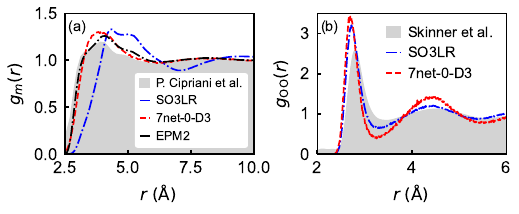}
    \caption{\label{fig:2}Radial distribution function (RDF) of \ce{CO2} and \ce{H2O} at liquid-like densities. (a) \ce{CO2} center-of-mass RDF $(g_m(r))$ sampled from MLMD simulation under canonical ensemble (NVT), $256$ \ce{CO2} in $3$ \unit{nm} box at $314$ \unit{K}, compared to the experimental reference data by P. Cipriani et al.(gray), EPM2 force field(black), SO3LR(blue), and 7net-0-D3(red). $g_m(r)=0.133g_\mathrm{CC}(r)+0.464g_\mathrm{CO}(r)+0.403g_\mathrm{OO}(r)$ (b) \ce{H2O} inter-oxygen RDF $(g_\mathrm{OO}(r))$ with $512$ molecules under canonical ensemble, compared to the reference data by Skinner et al. reference at $298$ \unit{K}(gray), SO3LR at $330$ \unit{K}(blue), and 7net-0-D3 at $330$ \unit{K}(red).}  
    \phantomsubcaption\label{fig:2a}
    \phantomsubcaption\label{fig:2b}
\end{figure}

Having benchmarked pairwise interaction energies in the low-density limit, we now assess whether these gas-phase trends persist in the condensed phase. Radial distribution functions (RDFs) serve as a direct structural probe, and we focus on SO3LR and 7net-0-D3 based on their comparatively better performance for \ce{CO2} and \ce{H2O} dimer energetics (\cref{fig:1}). \cref{fig:2a} shows the center-of-mass RDF for liquid \ce{CO2} obtained from $1~\unit{ns}$ molecular dynamics simulations of 256 molecules under canonical (NVT) ensemble at $314$ \unit{K} in a cubic cell with $3$ \unit{nm} box length at a density of $1.5\rho_c$ ($\rho_c = 10.6$ mol/L). The results are compared with neutron diffraction results reported by P. Cipriani et al.\cite{cipriani1997neutron} and with the modified extended primitive (EPM2) classical force field.\cite{harris1995carbon} EPM2 reproduces the first-peak position and height of reference data with the highest fidelity among all models tested. 7net-0-D3 also captures the first peak reasonably well, with only minor deviations in the intermediate region between $5$ and $8$ \unit{\AA}. In contrast, SO3LR shifts the first peak to approximately $4.2$ \unit{\AA}, overestimating the equilibrium intermolecular separation by roughly $0.3$ \unit{\AA}. In \cref{fig:1b}, SO3LR yields pairwise interaction energies in closer agreement with the reference than 7net-0-D3. By contrast, for the condensed-phase RDF (\cref{fig:2a}), 7net-0-D3 shows better consistency with the reference data than SO3LR. 

This reversal in model ranking for the same molecule suggests that the dominant factor governing model fidelity is not the molecular species itself but rather the thermodynamic environment, specifically the density, in which the molecules reside. Moreover, the pairwise interaction energy calculation can be viewed as probing the extremely low-density limit, whereas the condensed-phase RDF corresponds to a high-density environment. These observations indicate that the MLIPs examined here may exhibit a systematic density dependence in their accuracy. 

\cref{fig:2b} presents the oxygen-oxygen RDF for liquid water, computed from NVT simulations of $512$ molecules at $330$ \unit{K} and a density of $1.00$ \unit{g/cm^3}, compared with X-ray diffraction data from Skinner et al.\cite{skinner2013benchmark} Both SO3LR and 7net-0-D3 reproduce the first-peak position within $0.02$ \unit{\AA} of the reference value and capture the overall shape of the distribution through the second coordination shell. The agreement of both models with reference data for water, in contrast to the divergence observed for \ce{CO2}, is consistent with the training data composition discussed above. 

\subsection{\label{sec:2}Density-Dependent Transferability}

\begin{figure}
    \centering
    \includegraphics[width=\columnwidth]{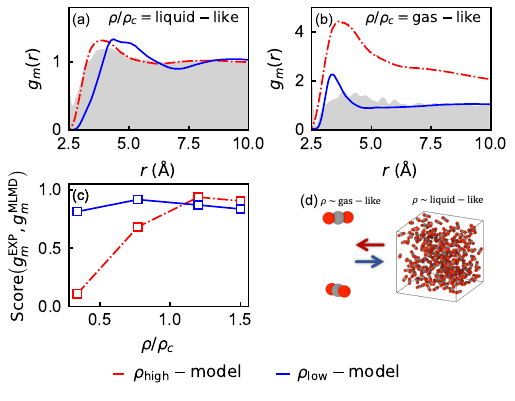}
    \caption{\label{fig:3}Density-dependent performance of MLIPs for \ce{CO2}. $g_m(r)$ at $310$ \unit{K} for (a) liquid-like density $\left( \rho/\rho_c = 1.50 \right)$ and (b) gas-like density $\left( \rho/\rho_c = 0.34 \right)$, where $\rho_c = 10.6\;\unit{mol/L}$. Gray shaded regions show experimental data, blue dash-dot lines show $\rho_\mathrm{low}-\mathrm{model}$, SO3LR, and red dashed lines show $\rho_\mathrm{high}-\mathrm{model}$, 7net-0-D3. (c) RDF similarity scores comparing simulation with experimental data as a function of density. (d) Schematic illustration of \ce{CO2} configurations at gas-like and liquid-like densities.}
    \phantomsubcaption\label{fig:3a}
    \phantomsubcaption\label{fig:3b}
    \phantomsubcaption\label{fig:3c}
    \phantomsubcaption\label{fig:3d}
\end{figure}

In this work, we define thermodynamic transferability as the ability of an MLIP to accurately describe systems across a broad range of thermodynamic state points, spanning
temperature and density conditions, without retraining. The contrasting performance of 7net-0-D3 and SO3LR for liquid \ce{CO2} (\cref{fig:1b,fig:2a}) suggests that model accuracy is governed not by the intrinsic interaction complexity of the molecule but by the density regime represented in the training data. To test this hypothesis, we systematically evaluated both models across a range of \ce{CO2} densities spanning gas-like to liquid-like conditions. To clarify the role of training data density, we hereafter denote 7net-0-D3 as $\rho_\mathrm{high}$-model and SO3LR  as $\rho_\mathrm{low}$-model, reflecting the dominant density regimes of their respective training databases. All simulations were performed at $310$ \unit{K} and compared with experimental reference data from prior neutron diffraction measurements.\cite{cipriani1997neutron}

\cref{fig:3a} shows the center-of-mass RDF at liquid-like density ($\rho/\rho_c = 1.50$, where $\rho_c = 10.6$ \unit{mol/L}). Consistent with \cref{fig:2a}, the $\rho_\mathrm{high}$-model reproduces the first peak of reference position and height, while the $\rho_\mathrm{low}$-model exhibits a systematic rightward shift of the first peak, indicating overestimated intermolecular separations. At gas-like density ($\rho/\rho_c = 0.34$, \cref{fig:3b}), the trend reverses. The $\rho_\mathrm{low}$-model captures the broad, featureless structure expected at low density, whereas the $\rho_\mathrm{high}$-model produces a pronounced first peak with a height approximately four times the bulk value. This artificial over-structuring indicates that the $\rho_\mathrm{high}$-model overestimates intermolecular attractions at low densities, driving highly local condensation.

To quantify this density-dependent behavior, we introduce an RDF similarity score defined as

\begin{equation}
  \mathrm{Score} = 1 - \frac{\int_{0}^{L} \left| g^\mathrm{TEST}(r) - g^\mathrm{VALID}(r)\right| \mathrm{d}r}{\int_{0}^{L} \left| g^\mathrm{TEST}(r) - 1.0 \right| \mathrm{d}r + \int_{0}^{L} \left| g^\mathrm{VALID}(r) - 1.0 \right| \mathrm{d}r}
  \label{eq:1}
\end{equation}

where $g^\mathrm{TEST}(r)$ is the RDF obtained from the simulation under evaluation and $g^\mathrm{VALID}(r)$ is that from the validation reference. The denominator sums each function's deviation from the ideal gas limit $(g(r)=1)$, which quantifies the total degree of structural ordering present in both distributions. The numerator integrates the absolute difference between the two RDFs, in which the ideal gas contribution cancels upon subtraction. Normalizing by the denominator ensures that the score measures the fractional reproduction of structural features rather than their absolute magnitude. A score of $1.0$ indicates perfect agreement and $0.0$ indicates complete disagreement.

\cref{fig:3c} presents the score as a function of reduced density $\rho/\rho_c$ for both models; the underlying RDFs at intermediate densities are provided in Fig.~S3 of the Supplementary Material. Here, $g^\mathrm{TEST}$ corresponds to $g_m^\mathrm{MLIP}$ obtained from each foundation model, and $g^\mathrm{VALID}$ to the experimental neutron diffraction data reported by P. Cipriani et al.\cite{cipriani1997neutron} The $\rho_\mathrm{high}$-model maintains high scores $(>0.9)$ at $\rho/\rho_c \geq 1.0$ but degrades as density decreases, falling below $0.1$ at $\rho/\rho_c = 0.34$. The $\rho_\mathrm{low}$-model displays the opposite trend, achieving its highest scores at low densities and declining toward liquid-like conditions.

This complementary behavior maps directly onto the density distributions of the underlying training databases (\cref{fig:3d}). The MPTrj database used for 7net-0 consists predominantly of solid-state configurations that sample high local densities and large coordination numbers, providing the $\rho_\mathrm{high}$-model with adequate coverage of liquid-like environments but insufficient representation of dilute, gas-like states. The MD DB used for SO3LR comprises gas-phase molecules and molecular fragments, equipping the $\rho_\mathrm{low}$-model with accurate descriptions of low-density configurations while limiting its extrapolation to condensed phases. These results establish that current foundation MLIPs do not reliably extrapolate outside the density range covered by their training data. Accurate prediction across multiple thermodynamic phases requires training databases whose density coverage encompasses the target conditions.

\subsection{\label{sec:3}Impact of Training Data Diversity}

\begin{figure}
    \centering
    \includegraphics[width=\columnwidth]{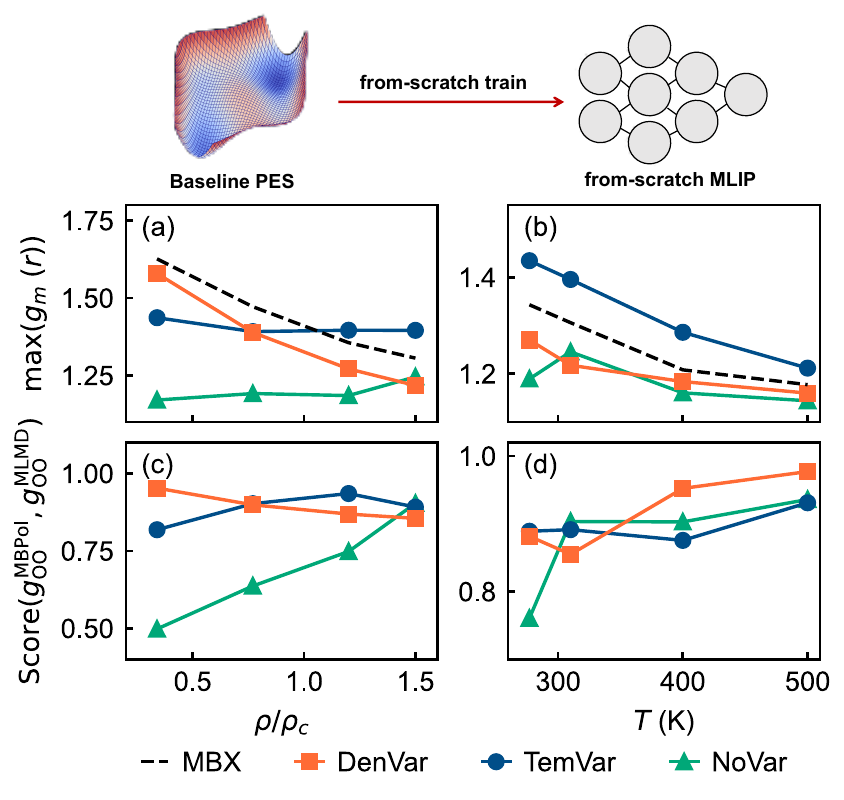}
    \caption{\label{fig:4}Effect of training-data diversity on the thermodynamic coverage of from-scratch \ce{CO2} MLIPs. Molecular dynamics simulations were performed for $256$ \ce{CO2} molecules in the NVT ensemble. The left column demonstrates the density variance system. Right column denotes the temperature variance system. Red color denotes DenVar dataset ($T = 310~\unit{K}$ and $\rho/\rho_c = 0.34, 1.50$), blue color denotes TempVar dataset ($T = 310, 500~\unit{K}$ and $\rho/\rho_c = 1.50$), and green color denotes NoVar dataset ($T = 310$ \unit{K} and $\rho/\rho_c = 1.50$). First-peak height of the center-of-mass RDF compared between MB-pol reference and each training strategy with (a) density variance and (b) temperature variance. RDF similarity score (\cref{eq:1}) with (c) density variance and (d) temperature variance with red square (DenVar), blue square (TempVar), and green triangle (NoVar).}
    \phantomsubcaption\label{fig:4a}
    \phantomsubcaption\label{fig:4b}
    \phantomsubcaption\label{fig:4c}
    \phantomsubcaption\label{fig:4d}
\end{figure}

As demonstrated by the \ce{H2O} results (\cref{fig:1c,fig:2b}), augmenting the training data with configurations that broaden the sampled density range should alleviate the density dependence of model accuracy. The density-dependent limitations of foundation MLIPs identified in \cref{fig:3} raise a practical question: Within a fixed training budget, what is the most effective strategy for expanding thermodynamic coverage? To address this, we trained from-scratch MLIPs for \ce{CO2} using the 7net architecture on training data generated from many-body polarizable (MB-pol)\cite{riera2020data, yue2022transferability} molecular dynamics simulations, eliminating uncertainties associated with heterogeneous database composition and exchange-correlation functional variations.

The baseline training set was constructed from a $1~\unit{ns}$ MB-pol trajectory of $64$ \ce{CO2} molecules at liquid-like density ($\rho/\rho_c = 1.50$, where $\rho_c = 10.6 \unit{mol/L}$) and $T = 310$ \unit{K} under NVT ensemble conditions, from which $1024$ configurations were randomly selected at $0.1$ \unit{ps} intervals. To assess the effect of thermodynamic diversity, we augmented this baseline with 1024 additional configurations according to three strategies, keeping the total training dataset size constant across all comparisons. NoVar adds $1024$ frames drawn from the same baseline condition ($\rho/\rho_c = 1.50$, $T = 310$ \unit{K}), increasing statistical sampling without introducing thermodynamic diversity. TempVar adds $1024$ frames sampled at the baseline density ($\rho/\rho_c = 1.50$) at $500$ \unit{K}. DenVar adds $1024$ frames sampled at the baseline temperature ($T = 310$ \unit{K}) at $\rho/\rho_c = 0.34$.

\cref{fig:4a} shows the first peak height of the center-of-mass RDF as a function of reduced density, with the corresponding full RDF profiles across temperature and density given in Figs.~S4 and~S5 of the Supplementary Material. The MB-pol reference exhibits a monotonic decrease in peak height with decreasing density, reflecting the progressive loss of short-range order as the system transitions from liquid-like to gas-like conditions. DenVar reproduces this trend, with peak heights that track the MB-pol reference across the entire density range. In contrast, NoVar and TempVar yield peak heights that remain nearly constant with density, failing to capture the structural evolution upon dilution. This behavior mirrors the $\rho_\mathrm{high}$-model results in \cref{fig:3}, confirming that models trained without density diversity cannot extrapolate beyond their training density. The corresponding RDF similarity scores (\cref{fig:4c}) quantify this difference. DenVar maintains higher scores across all densities, whereas NoVar and TempVar show progressive degradation as the density departs from the baseline condition.

The temperature-dependent results present a different picture. \cref{fig:4b} shows that all three strategies capture the decreasing peak height with increasing temperature, indicating that temperature extrapolation is comparatively less demanding. Among the three, DenVar achieves the highest peak-height agreement and RDF similarity scores at elevated temperatures (\cref{fig:4d}), outperforming even TempVar despite not explicitly including high-temperature configurations. This can be understood by recognizing that the effect of increasing temperature on liquid structure is mediated largely through changes in the effective local density sampled by molecular pairs. We substantiate this interpretation quantitatively in \cref{sec:5} through coordination number analysis. Density diversification therefore provides a more direct route to expanding the range of local coordination environments in the training set than temperature diversification alone.

These results establish that, within a fixed training budget, diversifying the density of the training configurations is the most effective strategy for broadening the thermodynamic coverage of from-scratch MLIPs. Density variation simultaneously improves transferability across both density and temperature axes, whereas temperature variation alone cannot compensate for missing density regimes. This finding reinforces the conclusion drawn from \cref{fig:3}: the density distribution of the training data is the primary factor governing thermodynamic transferability. Thermodynamic coverage cannot be guaranteed when the training budget is confined to a single density.

\subsection{\label{sec:4}Distillation for Robust Thermodynamic Coverage}

\begin{figure}
    \centering
    \includegraphics[width=\columnwidth]{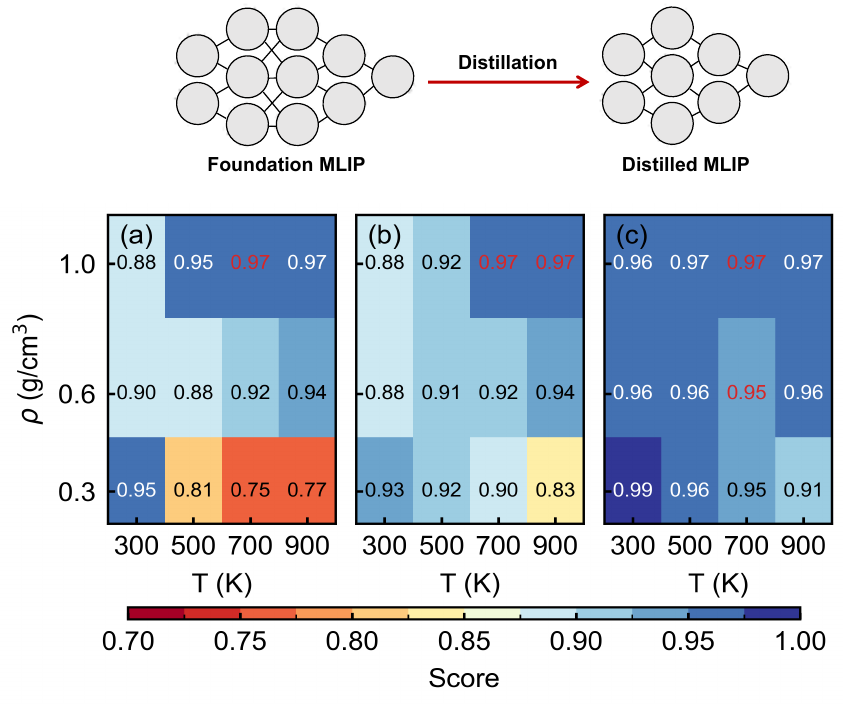}
    \caption{Thermodynamic coverage map of distilled MLIP with three different training strategies. (Top) Schematic of the distillation workflow. Configurations are sampled via molecular dynamics using a foundation MLIP and labeled with the same foundation model without additional DFT calculations. (Bottom) RDF similarity score maps across temperature ($300-900$ \unit{K}) and density ($0.3-1.0$ \unit{g/cm^3}) conditions. Red numbers indicate training conditions. (a) NoVar, single training condition ($T = 700$ \unit{K}, $\rho = 1.0\;\unit{g/cm^3}$). (b) TempVar, temperature-diverse training ($T = 700$ and $900$ \unit{K}, $\rho = 1.0\;\unit{g/cm^3}$). (c) DenVar, density-diverse training ($T = 700$ \unit{K}, $\rho = 0.6$ and $1.0 \unit{g/cm^3}$).}
    \label{fig:5}
    \phantomsubcaption\label{fig:5a}
    \phantomsubcaption\label{fig:5b}
    \phantomsubcaption\label{fig:5c}
\end{figure}

We next examined whether density-diversified training data retains its advantage in a distillation setting, where a computationally expensive foundation model is compressed into a lighter surrogate without additional first-principles labeling. The teacher model was the representational foundation model 7net-omni ($l_\mathrm{max} = 3$, $n_\mathrm{layer} = 5$)~\cite{kim_optimizing_2025}, and the student model adopted the 7net architecture ($l_\mathrm{max} = 1$, $n_\mathrm{layer} = 2$).  Because the inference cost scales with $l_\mathrm{max}$ and $n_\mathrm{layer}$, reducing these parameters lowers the computational complexity of the distilled model. Training data were generated by running NVT molecular dynamics simulations of $64$ \ce{H2O} molecules for $1$ \unit{ns} with the teacher model, from which 512 frames were randomly sampled. The energy, force, position, and atomic species labels produced by the teacher were used directly, with no further DFT recalculation. Following the protocol established in \cref{fig:4}, three training sets of equal size were constructed: NoVar samples only from the baseline condition, TempVar introduces temperature variation at fixed density, and DenVar introduces density variation at fixed temperature. The specific thermodynamic conditions used for each strategy are indicated by red-highlighted cells in \cref{fig:5}.

\cref{fig:5} presents RDF similarity score maps as a function of temperature ($300$-$900$ \unit{K}) and density ($0.3$-$1.0$ \unit{g/cm^3}) for the three distilled models, with the underlying RDFs at each thermodynamic state shown in Figs.~S6--S8 of the Supplementary Material. The DenVar model (\cref{fig:5c}) achieves scores above $0.91$ across all temperature and density conditions tested, demonstrating robust thermodynamic transferability. In particular, the DenVar model accurately reproduces liquid water structure at $300$ \unit{K} despite being trained on supercritical-phase configurations, confirming that density diversification enables extrapolation to thermodynamic states absent from the training set. The TempVar model (\cref{fig:5b}) shows moderate degradation at low densities, with scores declining to the $0.83-0.93$ range, and this degradation is amplified at lower temperatures. The NoVar model (\cref{fig:5a}) exhibits the most pronounced limitations, with the score dropping to $0.75$ at $\rho = 0.3$ \unit{g/cm^3} and $T = 700$ \unit{K}. The progressive improvement from NoVar to TempVar to DenVar is consistent across the entire thermodynamic plane, reproducing the hierarchy established for from-scratch training in \cref{fig:4}.

These distillation results reinforce two conclusions. First, density diversification of the training data yields the broadest thermodynamic coverage regardless of the training paradigm, whether from-scratch training on reference PES or distillation from a foundation model. Second, distillation combined with density-diverse sampling provides a practical route to developing lightweight MLIPs that maintain high structural fidelity across a wide range of thermodynamic conditions. The physical mechanism underlying this hierarchy is examined in the following section.

\subsection{\label{sec:5}Physical Origin of Density-Governed Transferability}

\begin{figure}
    \centering
    \includegraphics[width=\columnwidth]{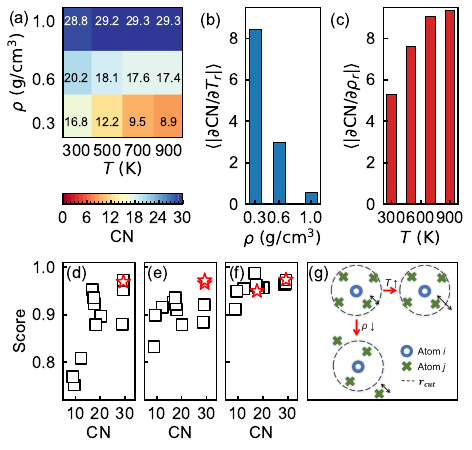}
    \caption{Physical origin of density-governed transferability. (a)~Reference coordination number $CN(r_{\mathrm{cut}})$ at $r_{\mathrm{cut}} = 6.0$~\AA\ as a function of temperature and density, computed from molecular dynamics of $64$~\ce{H2O} molecules using the 7net-omni teacher model. Numbers indicate the CN value at each condition. (b)~Mean absolute sensitivity of CN to reduced temperature, $\langle|\partial CN/\partial T_r |\rangle$, averaged over the temperature axis at each density. (c)~Mean absolute sensitivity of CN to reduced density, $\langle|\partial CN/\partial \rho_r |\rangle$, averaged over the density axis at each temperature. Reduced variables are defined as $T_r = T/T_{c}$ and $\rho_r = \rho/\rho_c$, with $T_c = 647$~\unit{K} and $\rho_c = 0.322$~\unit{g/cm^3}. (d--f)~RDF similarity score (\cref{eq:1}) plotted against reference CN for distilled models trained with (d)~NoVar, (e)~TempVar, and (f)~DenVar strategies. Open squares denote test conditions and red stars indicate training conditions. (g) Schematic illustration of local coordination environments within the model cutoff radius $r_{\mathrm{cut}}$ (dashed circles) as increasing temperature (horizontal red arrow) and decreasing density (vertical red arrow).}
    \label{fig:6}
    \phantomsubcaption\label{fig:6a}
    \phantomsubcaption\label{fig:6b}
    \phantomsubcaption\label{fig:6c}
    \phantomsubcaption\label{fig:6d}
    \phantomsubcaption\label{fig:6e}
    \phantomsubcaption\label{fig:6f}
    \phantomsubcaption\label{fig:6g}
\end{figure}

Having established that density diversity is the most effective strategy for achieving thermodynamic transferability, we now elucidate the underlying physical mechanism through coordination number (CN) analysis.

\cref{fig:6a} presents the reference coordination number as a heatmap across the temperature--density plane, where CN is evaluated at the model cutoff radius $(r_{\mathrm{cut}} = 6.0)$~\unit{\AA}. The heatmap reveals a pronounced anisotropy. CN varies sharply along the density axis, ranging from approximately $9$ at low density ($\rho = 0.3$~\unit{g/cm^3}, $900$~\unit{K}) to $29$ at high density ($\rho = 1.0$~\unit{g/cm^3}), whereas it remains nearly constant along the temperature axis at high density ($28.8-29.3$ across the full temperature range at $\rho = 1.0$~\unit{g/cm^3}). At low density, temperature exerts a more appreciable effect (CN decreasing from $16.8$ to $8.9$ at $\rho = 0.3$~\unit{g/cm^3}), yet this variation is itself a consequence of temperature-induced changes in the effective local density sampled by molecular pairs.

\cref{fig:6b,fig:6c} quantify this anisotropy sensitivity using reduced variables $T_r = T/T_c$ and $\rho_r = \rho/\rho_c$ ($T_c = 647$~\unit{K}, $\rho_c = 0.322$~\unit{g/cm^3}). The sensitivity of CN to reduced temperature (\cref{fig:6b}) decreases sharply with increasing density, falling from approximately $8.44$ at $\rho = 0.3$~\unit{g/cm^3} to $0.57$ at $\rho = 1.0$~\unit{g/cm^3}. In contrast, the sensitivity of CN to reduced density (\cref{fig:6c}) ranges from $5.31$ to $9.36$ across all temperatures, exceeding the temperature sensitivity. 

This implies that when temperature variation (TempVar) is applied, the resulting CN coverage depends strongly on the choice of density at which the training data are sampled. By contrast, density variation (DenVar) exhibits comparatively weak dependence on temperature, suggesting that a training set constructed with density diversity yields a model with broader CN coverage regardless of the temperature conditions selected for data generation.

\cref{fig:6d,fig:6e,fig:6f} directly connect this CN analysis to model transferability. Each panel plots the RDF similarity score against the reference CN for models trained with the NoVar, TempVar, and DenVar strategies, respectively. For NoVar (\cref{fig:6d}), high scores are confined to the CN range near the training condition (CN~$\approx 29$), with performance degrading sharply as CN decreases (score~$\approx 0.76$ at CN~$\approx 9$). TempVar (\cref{fig:6e}) exhibits a similar pattern, because varying temperature at fixed density fails to expand the CN range of the training data. In contrast, DenVar (\cref{fig:6f}) maintains scores above $0.91$ across the entire CN range from $9$ to $29$. The training conditions of DenVar (stars) span CN $18$ and 29, providing the model with exposure to a broad range of coordination environments. These results indicate that the coordination number serves as the principal physical descriptor governing the thermodynamic coverage of MLIPs.

The monotonic relationship between CN and the RDF similarity score observed in \cref{fig:6d} illustrates the performance degradation that arises when a model is trained on configurations sampled at a single CN value and evaluated outside its training range. \cref{fig:6e} shows a qualitatively similar degradation: TempVar introduces two training conditions whose CN values are nearly indistinguishable, limiting the model's ability to resolve structurally distinct environments. In contrast, DenVar (\cref{fig:6f}) supplies training configurations at well-separated CN values, enabling CN to act as a more discriminating feature and yielding uniformly high scores across the full thermodynamic plane. These trends provide a critical clue that model performance depends on the CN diversity encoded in the training data.

The relationship between CN and the two thermodynamic variables can be understood from how each variable affects the local coordination environment. As illustrated schematically in \cref{fig:6g}, temperature controls the kinetic-energy distribution and affects CN only indirectly through structural fluctuations. Varying temperature therefore modulates this variance and alters the mean CN only indirectly, through fluctuations in the local density sampled by molecular pairs. Varying density, by contrast, directly changes the mean interparticle separation and thus the number of neighbors within the cutoff radius, providing a more effective route to shifting the mean CN. This interpretation is consistent with previous studies on simple and complex fluids, which established that the coordination number undergoes sharp transitions along the density axis, serving as a topological descriptor that distinguishes qualitatively different fluid regimes \cite{yoon2018topological,yoon2019topological,yoon2020topological}.

The connection between CN and MLIP accuracy follows directly from the local-energy decomposition inherent to modern MLIP architectures. Modern MLIPs rest on the assumption that the total energy can be decomposed into a sum of local atomic contributions, $E = \sum_i E_i$.~\cite{behler2007generalized, zeng2025deepmd, batzner20223, Batatia2022mace, park_scalable_2024} Each local energy $(E_i)$ is a learned function of the topology and geometry of the neighbor set within a cutoff radius, so that $E_i$ depends implicitly on the CN. Density variation directly alters this coordination topology, and because it does so more efficiently than temperature variation, it exerts a more critical influence on MLIP training. It is precisely in fluidic systems, where CN spans a wide range across thermodynamic states, that density-diverse training most effectively broadens thermodynamic coverage.

\section{Conclusions}

This study demonstrates that the density distribution of training data is the primary factor governing the thermodynamic transferability of MLIPs in fluid-phase systems relevant to chemical engineering. Thermodynamic transferability is a direct consequence of training data composition, not an inherent property of model architecture, and prioritizing density diversity in training set construction offers the most resource-efficient path to MLIPs that maintain structural fidelity across the gas-, liquid-, and supercritical regimes encountered in process operation.

Evaluation of representative foundation MLIPs revealed that models extrapolate poorly beyond the density regime of their training data. The complementary failure modes of the solid-state-trained 7net-0 and the gas-phase-trained SO3LR for \ce{CO2}, with the former producing unphysical clustering at gas-like densities and the latter overestimating intermolecular separations in the liquid phase, confirmed that the density coverage of the training database, rather than model architecture, sets the transferability boundary.

Controlled from-scratch training and distillation experiments under fixed computational budgets established that density diversification is the most effective strategy for broadening thermodynamic coverage. Density-diverse training achieved the highest structural fidelity across both density and temperature axes simultaneously, outperforming temperature diversification even for temperature-dependent properties. This hierarchy was reproduced in distillation experiments on \ce{H2O}, where the density-diverse protocol yielded RDF similarity scores above $0.91$ across all tested conditions, confirming the generality of this finding across molecular systems and training paradigms.

The physical origin of this hierarchy is rooted in the local-energy decomposition inherent to MLIP architectures. Because each atomic energy $E_i$ is determined by the neighbors within a finite cutoff radius, the coordination number within this receptive field directly parametrizes the space of local environments that the model must learn to represent. Coordination number analysis showed that density variation alters this quantity far more effectively than temperature variation, which leaves the coordination landscape largely unchanged at fixed density and influences liquid structure only indirectly through local density fluctuations. Density diversification therefore expands the model's representational coverage along the very axis of local coordination on which its energy predictions depend.

These results carry direct implications for MLIP development. For model developers, incorporating low-density and gas-phase configurations alongside the solid-state data dominating current databases is essential for extending transferability to fluidic applications. For end users, the validation protocol presented here, combining pairwise energy profiles, density-dependent RDF similarity scores, and coordination number analysis, provides a practical framework for assessing whether a model's training distribution covers the target thermodynamic conditions. Beyond single-component molecular systems, extending the density-diversity principle to multicomponent mixtures and reactive systems, where compositional degrees of freedom introduce additional dimensions of thermodynamic variability, represents a natural next step toward universally transferable MLIPs.

\section{Methods}

\subsection{Pairwise Interaction Energy}

\begin{equation}
    E_{A\cdots A}(r) = E_{\mathrm{dimer}}(r) - 2\,E_{\mathrm{monomer}}
\end{equation}

The pairwise interaction energy $E_{A\cdots A}$ is defined as the difference between the dimer energy $E_{\mathrm{dimer}}$ at intermolecular separation $r$ and twice the monomer energy $E_{\mathrm{monomer}}$. All energies are computed for isolated systems without periodic boundary conditions. Each force field or potential was used to perform geometry optimization of the monomer. The resulting geometry was held fixed while the dimer interaction energy was evaluated as a function of $r$. All calculations were carried out within a unified framework using the Atomic Simulation Environment (ASE) Python package.\cite{hjorth2017atomic}

Reference interaction energies were computed with the Gaussian 16 program.\cite{g16} The exchange-correlation functional was PBE at the GGA level, augmented with Grimme's D3 dispersion correction with Becke--Johnson damping (D3(BJ)).\cite{perdew1996generalized,grimme2010consistent,grimme2011effect} The basis set was the second-generation Karlsruhe triple-zeta valence polarization set (def2-TZVP) of Weigend and Ahlrichs.\cite{weigend2005balanced}

\subsection{Radial Distribution Function and Coordination Number}

The radial distribution function (RDF) is defined as
\begin{equation}
  g(r) = \frac{1}{4\pi r^2 \rho \Delta r} 
  \left\langle n(r, r+\Delta r) \right\rangle
\end{equation}
where $\rho$ is the number density, $\Delta r$ is the bin width, and $\langle n(r, r+\Delta r) \rangle$ is the average number of particles in a spherical shell at distance $r$. A bin width of $\Delta r = 0.05$~\unit{\AA} was used. 

The coordination number is defined as
\begin{equation}
  CN(r) = 4\pi\rho \int_0^{r} g(r')\, r'^2 \,\mathrm{d}r'
\end{equation}
where $g(r')$ is the RDF. 

\subsection{Machine Learning Molecular Dynamics}

All molecular dynamics simulations were run for $1~\unit{ns}$ with a timestep of $0.5~\unit{fs}$. MLIP based molecular dynamics was performed using the ASE package. Temperature control under the NVT ensemble was achieved with a Nos\'{e}--Hoover chain thermostat\cite{martyna1999molecular, tuckerman2023statistical} with a damping coefficient of $50$ \unit{fs}, chain length of $3$, and a single thermostat sub-step per integration step. Dispersion interactions were treated with the D3(BJ) correction via the \texttt{torch-dftd} package\cite{takamoto2021pfp}, using PBE as the reference functional. All simulations were accelerated on GPU hardware. Initial configurations of \ce{H2O} and \ce{CO2} systems were generated with PACKMOL\cite{martinez2009packmol}.

\subsection{Training Database Generation for From-Scratch MLIP}

Training data for the from-scratch MLIP were generated using the MB-pol potential, implemented through the MBX\cite{riera2023mbx} code coupled to i-PI.\cite{ceriotti2014pi} A two-body cutoff of $9.0~\unit{\AA}$ and a three-body cutoff of $4.6~\unit{\AA}$ were employed. NVT dynamics used a Langevin thermostat with $\tau = 50~\unit{fs}$. A $1.0~\unit{ns}$ NVT trajectory of a $64$ molecule \ce{CO2} system was generated, and configurations were saved at $0.1~\unit{ps}$ intervals to reduce inter-frame correlation. Training structures were drawn by random sampling from this trajectory.

\subsection{Hyperparameters for From-Scratch MLIP}

The from-scratch model was trained using the SevenNet architecture\cite{park_scalable_2024} with the following hyperparameters. Cutoff radius $5.0$~\unit{\AA}, channel size $32$, maximum angular momentum $l_{\max} = 2$, three convolutional layers, and SE(3) equivariance. The Adam optimizer was used with an initial learning rate of $5 \times 10^{-3}$ and an exponential learning rate schedule with decay factor $\gamma = 0.99$. Force and stress loss weights were set to 0.1 and $10^{-6}$, respectively. Training ran for 1000 epochs with a batch size of $4$. $10\%$ of the data was reserved for validation.

\subsection{Hyperparameters for Distilled MLIP}

The distilled model used a cutoff radius of $6.0$~\unit{\AA}, channel size $32$, $l_{\max} = 1$, and two convolutional layers. All other training hyperparameters were identical to those of the from-scratch model. The $6.0$~\unit{\AA} cutoff matches that of the teacher model, SevenNet-omni.\cite{kim_optimizing_2025} The parameters $l_{\max}$ and the number of convolutional layers are the primary determinants of inference cost. Reducing those parameters from the teacher values ($l_{\max} = 3$, $n_{\mathrm{conv}} = 5$) to $(1,\,2)$ increases inference speed at the expense of some accuracy.

% \section*{Acknowledgments}
% This work was supported by the National Research Foundation of Korea (NRF) grant funded by the Korea government (MSIT)~(RS-2025-25424498), by the Technology Innovation Program(RS-2025-19532970) funded by the Ministry of Trade, Industry and Resources(MOTIR, Korea), and by the Ministry of Education of the Republic of Korea and the National Research Foundation of Korea(RS-2024-00343871).

% \section*{Author contributions}
% Conceptualization, M.K.; methodology, M.K., M.Y.H. and W.B.L.; investigation, M.K., S.K, J.Y.J. and W.B.L.; visualization, M.K., M.Y.H. and W.B.L.; writing-–original draft, M.K., M.Y.H. and W.B.L.; writing-–review \& editing, M.K., S.K, J.Y.J., M.Y.H. and W.B.L.; funding acquisition, W.B.L.; resources, M.Y.H. and W.B.L.; supervision, M.Y.H and W.B.L.

% \section*{Declaration of competing interest}
% The authors declare that they have no known competing financial interests or personal relationships that could have appeared to influence the work reported in this paper.

% \section*{Data availability}
% The data underlying the figures of this study are openly available at \url{https://github.com/minu928/MLFFTrans}.

\clearpage
\bibliography{references}

\end{document}